\documentclass[twocolumn,english,pra,showpacs,floatfix]{revtex4}
\usepackage[T1]{fontenc}
\usepackage[latin1]{inputenc}
\usepackage{graphicx}
\usepackage{amssymb}

\makeatletter

\usepackage{graphicx}

\usepackage{babel}
\makeatother

\newcommand \bea{\begin{eqnarray}}
\newcommand \eea{\end{eqnarray}}

\newcommand \la{\raisebox{-.5ex}{$\stackrel{<}{\sim}$}}
\newcommand{\av}[1]{\langle{#1}\rangle}

\begin{document}
\title{Induced interactions \& superfluidity
in optical lattices with multi-component Fermi gases}
\author{H. Heiselberg}
\affiliation{Applied Research, DALO, Lautrupbjerg 1-5, DK-2750 Ballerup, Denmark}
\begin{abstract}

Many-body effects on superfluidity and transition temperatures are
calculated for optical lattices and uniform systems with ultracold
multi-component Fermi gases. The induced interactions depend
sensitively on the interactions between the multi-components and their
densities.  The s- and d-wave pairing gaps and critical temperatures
are calculated for optical lattices at low (dilute) filling as well as
near half filling to leading orders in the interaction strength with
and without induced interactions. They can deviate strongly from
dilute two-component systems and affect the phase diagram.  In two
dimensional optical lattices the induced interactions are singular at
half filling and strongly affect s- and d-wave superfluidity.
\end{abstract}
\maketitle


Bardeen, Cooper, \& Schrieffer's theory of superconductivity (BCS)
was early on extended with the many-body effect of induced
interactions by Gorkov \& Melik-Bharkudarov \cite{Gorkov} who found
that the BCS pairing gap and critical temperature was reduced by a
factor $(4e)^{1/3}$ in uniform three dimensional (3D) systems. 
Reductions have been observed in
superconductors, $^3$He \cite{Leggett}, and in ultra-cold Fermi atom
systems in traps and optical lattices
\cite{Thomas,Chin,Jordens,Schneider}.  The induced interactions were
further discussed and extended to any number of spin
states as well as including bosons \cite{II}.  Recently, the induced
interactions were calculated numerically for lattices with
two-component fermions within the Hubbard model, and near
half filling they gaps and critical temperatures were strongly suppressed
\cite{Kim}.  Also induced interactions were
calculated for uniform 3D $^6$Li gases with three spin states
\cite{Martikainen}.

It has become experimentally feasible to vary the interactions
strength between states in atomic traps and optical lattices by using
Feshbach resonances and to study the crossover from BCS to molecular
BEC in mixtures of Fermi (and Bose) gases. Recently, multi-component systems of
Fermi atoms with several atomic states of internal hyperspin have been produced
such as $^6$Li \cite{Jochim}, $^{137}$Yb with six nuclear
spin states \cite{Kitagawa}, and heteronuclear mixtures of $^{40}$K and $^6$Li
\cite{Grimm}.
Such multi-component systems have intriguing similarities with neutron, nuclear
and quark matter where, for example, color superconductivity between the
2 spin, 8 color and 2-3 flavor states may occur \cite{Alford}.
Ultracold Fermi atoms provide a highly
controlled environment in which superfluidity can be studied in detail
over a wide range of multi-component densities, temperatures,
interaction strengths, etc. Additionally, 2D optical lattices 
have the important quality that by changing the on-site interaction from
attractive to repulsive the BCS s-wave pairing changes to d-wave
pairing, which is believed to be responsible for high temperature
superconductivity \cite{Anderson}.  2D optical lattices may therefore
also emulate high temperature superfluidity as well as Mott
insulators, anti-ferromagnetic and other interesting phases and
transitions \cite{Hofstetter,Bloch,gos}.

The purpose of this work is to calculate the s- and d-wave pairing gaps and
critical temperatures for multi-component Fermi gases
in optical lattices and uniform gases with and without induced
interactions. For this purpose we employ 
the Hubbard Hamiltonian on a D-dimensional lattice, which
describes optical lattices that are sufficiently deep for a one-band model
to apply,
\bea \label{Hubbard}
 H &=& \sum_{i,\sigma<\sigma'} U_{\sigma,\sigma'}
       \hat{n}_{i\sigma} \hat{n}_{i\sigma'}
-t\sum_{\av{ij},\sigma} \hat{a}_{i\sigma}^\dagger \hat{a}_{j\sigma} \,.
\eea
Here $\hat{a}_{i\sigma}^\dagger$ is the Fermi 
creation operator of the hyperspin states $\sigma=1,2,...,\nu$, 
$n_{i\sigma}=\hat{a}_{i\sigma}^\dagger \hat{a}_{i\sigma}$ the density
and $\av{ij}$ denotes nearest neighbours with hopping parameter $t$.

We shall investigate s-wave pairing between to spin states, e.g.
1 and 2, due to an
attractive on-site interaction $U_{1,2}<0$ in the presence of
multi-components $j=3,...,\nu$ with onsite interactions $U_{ij}$. 
The mean field gap equation for singlet superfluidity 
at zero temperature is \cite{Micnas,II} 
\bea \label{sgap}
 \Delta_{\bf p'} = -\frac{1}{M} \sum_{\bf p} U_{12}^{eff}({\bf p'},{\bf p})
    \Delta_{\bf p} \frac{\tanh(E_{\bf p}/2k_BT)}{2E_{\bf p}} \,,
\eea
where $M$ is the number of lattice points,
$E_{\bf p}=\sqrt{(\epsilon_{\bf p}-\mu_1)^2+\Delta_{\bf p}^2}$ with
$\epsilon_{\bf p}=2t\sum_{i=1,D}(1-\cos p_i)$, and the chemical potentials
are assumed equal $\mu_1=\mu_2$ for the pairing components.
 The effective interaction $U_{12}^{eff}=U_{12}+U_{ind}({\bf p'},{\bf p})$ 
includes induced interactions which will be calculated below.
The density 
$n=1-\sum_{\bf p}(\epsilon_{\bf p}/E_{\bf p})\tanh(E_{\bf p}/2k_BT)/M$, is
also the filling fraction in units where the lattice constant is unity.
Momenta are in the first Brillouin zone only $|p_i|\le\pi$.

Due to particle-hole symmetry results also apply replacing the density
by $(2-n)$ and chemical potentials by $(4Dt-\mu)$, where $4Dt$ is
the bandwidth in $D$ dimensions. At low filling $\epsilon_p=tp²$ and
results also apply to uniform systems replacing $t=1/2m$, where $m$ is
the particle mass.

We first calculate the zero temperature gaps for weak attraction
$|U_{12}|\ll t$ excluding induced interactions. The s-wave gap
$\Delta_{\bf p'}=\Delta_0$ is then momentum-independent because the
interaction $U_{12}$ is.  The gap equation (\ref{sgap}) reduces to: 
$1=-(U_0/M)\sum_{\bf p}1/2E_{\bf p}$, from
which the gap can be calculated analytically at low and half filling 
to leading orders in 3D and 2D (for 1D see, e.g. \cite{Marsiglio}).

{\bf The 3D lattice} has a critical
coupling $U_c=-M/(\sum_{\bf q} 1/2\epsilon_{\bf q})
=-8\sqrt{2}t/[\sum_{l=0}^\infty P_{2l}(\sqrt{9/8})(2l-1)!!/2^{2l}l!]
\simeq -7.913t$,
\cite{Svistunov} where a two-body bound state can be formed at zero
density ($\mu_1=0$). This threshold corresponds to the unitary limit of
infinite scattering length in the uniform system (see e.g. \cite{long}),
and naturally enters the sum in the gap equation.
The dilute gap becomes to leading orders
\bea \label{G3D}
  \Delta_0^{3D} = \frac{8}{e^2} \mu_1 \exp\left[\frac{4\pi^2 t}{k_F}
   \left(\frac{1}{U_{12}}-\frac{1}{U_c}\right) \right] ,\, n\ll1 ,
\eea
at low densities where $\mu_1=tk_{F,1}^2\ll t$.
The level density at the
Fermi surface $N(\mu_1)=k_{F,1}/(4\pi^2t)$ enters in the exponent as in standard
BCS theory. $U_c$ acts as a cutoff in
the one-band Hubbard model which in the uniform (continuum) limit 
is absorbed into the scattering lengths $a_{ij}$ between states $i$ and $j$.
Thus Eq. (\ref{G3D}) is the finite lattice equivalent of the
gap in the uniform system $\Delta_0^{3D}=(8/e^2)\mu_1\exp(\pi/2k_{F,1}a_{12})$,
with $U_{12}U_c/(U_c-U_{12})$ replaced by $4\pi a_{12}/m$ such that
threshold $U_{ij}=U_c$ corresponds to $|a_{ij}|=\infty$. 

Near half filling the level density is surprisingly constant
$N(\mu_1)\simeq 0.143/t$ in a wide range $4t\le\mu_1\le8t$ around half filling. 
The gap can therefore is to leading orders
\bea\label{G3D1}
  \Delta_0^{3D} = \alpha t \exp\left[\frac{1}{N(\mu_1)U_{12}}\right] ,\, n\simeq 1,
\eea
where the prefactor $\alpha\simeq6.544..$ is calculated numerically.

{\bf The 2D lattice} has a superfluid s-wave gap that be calculated
from the gap equation \cite{Dupuis}
\bea \label{G2D}
  \Delta_0^{2D} = \sqrt{8\pi\mu_1 t} \,\exp\left[\frac{4\pi t}{U_{12}}\right] \,,
  \, n\ll 1,
\eea
when the density $n=k_F^2/2\pi$ is small. 
The 2D level density is $N(\mu_1)=1/(4\pi t)$.
Eq. (\ref{G2D}) assumes that the
on-site coupling and the gap are small such that 
$\Delta_0^{2D}\la\mu_1$. This is not fulfilled at sufficiently
low densities where instead $\Delta_0^{2D}=4\pi t\exp(8\pi t/U_{12})$, 
which also is the two-body binding energy in 2D at zero density.
There is always a two-body bound state in 2D with purely attractive
interaction and therefore $U_c$ vanishes. In 3D a similar pair condensate
(a molecular BEC) 
appears when $a_{12}>0$ corresponding to $U_{12}<U_c$.

For intermediate fillings the pairing is
more complicated. Near half filling $n\simeq1$ the level density
$N(\epsilon)=\ln(16t/|\epsilon-4t|)/(2\pi^2t)$
has a logarithmic singularity due to the van Hove singularity.
Calculating the r.h.s. of the gap equation therefore gives a double log:
$1=|U_{12}|/(4\pi^2)\ln^2(32t/\Delta_0^{2D})$, to leading logarithmic orders,
resulting in the gap \cite{Dupuis}
\bea
 \Delta_0^{2D}= 32t \exp(-2\pi\sqrt{t/|U_{12}|}) \,,\, n\simeq 1.
\eea

{\bf Induced interactions} were originally calculated for dilute 3D 
two-component systems by Gorkov \&
Melik-Bharkudarov \cite{Gorkov} and extended to
multi-component Fermi gases also with Bose atoms
\cite{II}. The induced
interactions arise from two types of second order diagrams \cite{Gorkov,II,Martikainen}
\bea \label{ind}
  U_{ind}({\bf p'},{\bf p}) &=& -\frac{U_{12}^2}{M}
  \sum_{\bf q} \frac{f(\xi_1({\bf k'+q}))-f(\xi_1({\bf q}))} 
  {\xi_1({\bf k'+q})-\xi_1({\bf q})}  \nonumber\\
  &+& 
  \sum_{j,{\bf q}}\frac{U_{1j}U_{2j}}{M}
 \frac{f(\xi_3({\bf k+q}))-f(\xi_3({\bf q}))}
  {\xi_3({\bf k+q})-\xi_3({\bf q})}  .
\eea
$f$ is the Fermi distribution of $\xi_j({\bf q})=\epsilon_{\bf q}-\mu_j$,
${\bf k'}={\bf p}+{\bf p'}$ and ${\bf k}={\bf p}-{\bf p'}$,
where ${\bf p}_i$ are the momenta of the two pairing spin states. 
Induced interactions due to particle-hole loop diagrams 
from $j=3,...,\nu$ spin states are
responsible for the second sum in Eq. (\ref{ind}) and has opposite sign.
Near half filling
the induced interactions are enhanced considerably
because Umklapp processes enhance the
phase space around where the denominator of Eq. (\ref{ind}) is
singular. 

For later convenience we write the induced interactions
\bea \label{Uind}
U_{ind}=U_{12}^2N(\mu_1)\chi(\omega_1)-
  \sum_{j=3}^\nu U_{1j}U_{2j}N(\mu_j)\chi(\omega_j) ,
\eea
in terms of a screening function $\chi$ of
$\omega_1={\bf k'}/2k_{F,1}$ and $\omega_j={\bf k}/2k_{F,j}$.
At low density the screening function reduces in 3D to the Lindhard function
$\chi^{3D}(\omega)=\frac{1}{2}+
       \frac{1-\omega^2}{4\omega} \ln\left|\frac{1+\omega}{1-\omega}\right|$,
whereas in 2D 
\bea \label{2DL}
\chi^{2D}(\omega) = \left\{\begin{array}{cll}
  1&,& \omega<1 \\  1-\sqrt{1-1/\omega^2}&,& \omega>1  
\end{array} \right\}.
\eea
For weak interactions the gap is small and pairing occurs near the Fermi surface
only. To calculate the s-wave gap we can therefore replace
$\chi(\omega)$ in Eq. (\ref{ind}) by its  momentum
average $\bar{\chi}(\mu_1,\mu_j)$ 
over the Fermi surface $|\epsilon_{{\bf p}}|=|\epsilon_{{\bf p'}}|=\mu_1$
\cite{Gorkov,II}.
At low densities $\mu_j=tk_{F,j}^2$ and 
$\omega_j=(k_{F,1}/k_{F,j})\sqrt{(1+\cos\theta)/2}$, where $\theta$
is the angle between ${\bf p}$ and ${\bf p'}$.
Averaging over this angle in Eq. (\ref{2DL}) gives 
\bea
  \bar{\chi}^{2D}(\mu_1,\mu_j)= \left\{\begin{array}{cll}
   1&,&\mu_j\ge \mu_1  \\ k_{F,j}/k_{F,1} &,& \mu_j\le \mu_1 
\end{array} \right\}.
\eea
The screening function $\bar{\chi}^{2D}$ increases with multi-component filling
only up to equal filling where a plateau appears.
The result $\bar{\chi}^{2D}(\mu_1,\mu_1)=1$ was earlier 
found by Nishida \& Son \cite{Nishida} for uniform systems. In 
3D uniform systems $\bar{\chi}^{3D}=1$ when $\mu_j\gg\mu_1$, 
$\bar{\chi}^{3D}=(1/3)\log(4e)$ when $\mu_j=\mu_1$ and
$\bar{\chi}^{3D}=(\mu_j/3\mu_1)\log(\mu_1/\mu_j)$ 
when $\mu_j\ll\mu_1$. \cite{Martikainen}.

\begin{figure}
\includegraphics[scale=0.46,angle=0]{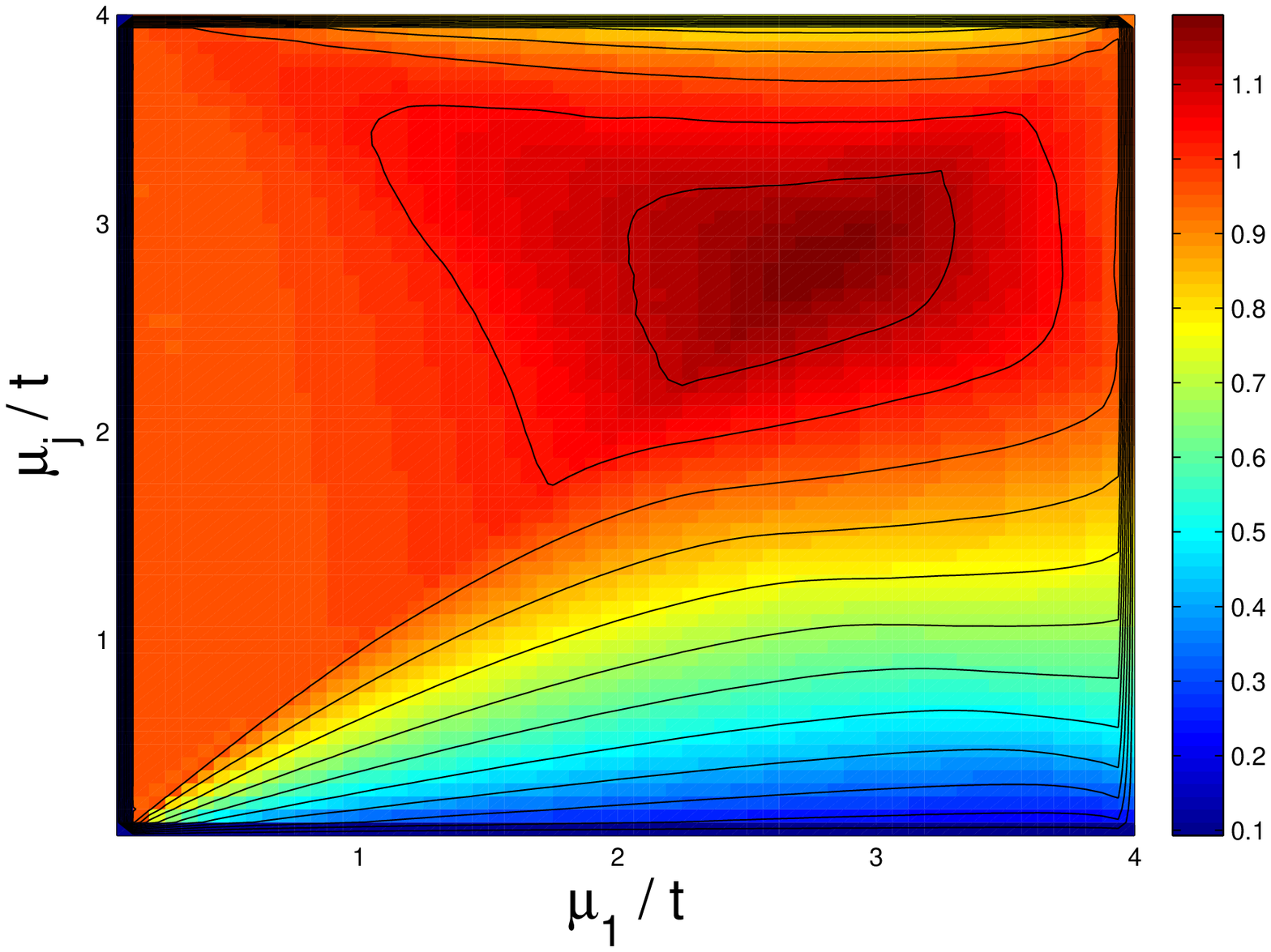}
\vspace{-0.5cm}
\caption{Averaged screening function $\bar{\chi}^{2D}(\mu_1,\mu_j)$} in a 2D optical
lattice vs. multi-component chemical potentials.
\end{figure}

$\bar{\chi}^{2D}(\mu_1,\mu_j)$ is shown in Fig. 1 for general filling.
The plateau $\bar{\chi}^{2D}\simeq1$ extends almost to half filling.
Due to particle-hole symmetry the screening function is symmetric when
any chemical potential $\mu$ is replaced by $4Dt-\mu$. 
We find that the induced interactions has three logarithmic singularities at 
$\mu_1=0,4t,8t$ all at $\mu_j=4t$.
The screening function itself is finite at these points.
The singularity at half fillings
was calculated numerically by Kim et al. \cite{Kim} for a two component
system. It can be calculated analytically by observing
that near half filling the Fermi surface of components 1 and 2 
is a square in 2D and the momenta ${\bf p}$ and ${\bf p'}$ thus move on
either parallel or perpendicular Fermi surface sections. 
In the parallel case, defining $p_{\pm}=(p_{x}\pm p_{y})/2$, $p'_{\pm}=(p'_{x}\pm p'_{y})/2$,
we notice that either $k_+=p_{+}+p'_{+}$ or $k_-=p_{-}+p_{-}'$ 
vanishes, and the sum in Eq. (\ref{ind}) (taking, e.g., $k_+=0$)
therefore reduces to 
$\chi^{2D}(\omega) \simeq (4tN(\mu_j))^{-1}
  \sum_{\bf q} f(\xi_3({\bf q}))/(\cos q_+[\cos(q_-+k_-)-\cos q_-])$.
Near half filling $\delta=4-\mu_j/t$ is small and
the momenta effectively extend to
$|q_-|\le\pi/2$ and $|q_+|\le \pi/2-\delta$,
and the remaining integrals as well as the
averaging of $k_-$ over the square Fermi surface can be calculated. 
Summing over $q_+$ gives a logarithmic singularity proportional to
the level density so that $\bar{\chi}^{2D}(4t,\mu_j)\simeq \pi/4$ as
$\mu_j\to 4t$.

Including the induced interactions $U^{eff}=U_{12}+\bar{U}_{ind}$
leads to a gaps simply replacing $U_{12}$ by $U^{eff}$
in Eqs. (3-6). When the induced interactions are small compared to the 
on-site interactions, which is the case for weak interactions except
near half filling in 2D, we obtain
\bea \label{D}
  \Delta = \Delta_0 \exp\left[ -\bar{\chi}(\mu_1,\mu_1) +
 \sum_{j=3}^\nu \frac{U_{1j}U_{2j}}{U_{12}^2}
\frac{N(\mu_j)}{N(\mu_1)}   \bar{\chi}(\mu_1,\mu_j)  \right]
\eea
where $\Delta_0$ is the uncorrected gap as e.g. given in Eqs. (3-5).
In a two-component systems 
the suppression factor $e^{-\bar{\chi}}$ is
consistent with previous estimates of many-body effects \cite{manybody,Kim}.

In an ultracold $^6$Li,
$^{40}$K or heteronuclear gases the many hyperfine states generally have
different scattering lengths depending on magnetic field. Varying the
lattice depth $V_0$ and wavenumber $k$ allows further tuning of the resulting
onsite couplings $U_{ij}=E_Ra_{ij}k\sqrt{8/\pi}\xi^3$ and
hopping $t=E_R(2/\sqrt{\pi})\xi^3e^{-2\xi^2}$, where $E_R=k^2/2m$ is the recoil
energy and $\xi=(V_0/E_R)^{1/4}$,
and subsequently also the effective interactions included in Eq. (\ref{D}).

In gases with ultracold $^{173}$Yb the scattering lengths between the
$\nu=6$ nuclear spin states are the same. \cite{Kitagawa}
If we assume equal population of all states
we find from Eq. (\ref{D}) that the induced interactions change the gap
at low densities by a factor
\bea
\Delta^{2D}=\Delta^{2D}_0e^{\nu-3}  \,,
\eea
in 2D. In 3D $\Delta^{3D}=\Delta^{3D}_0(4e)^{\nu/3-1}$, since
$\bar{\chi}^{3D}(\mu_1,\mu_1)=\ln(4e)/3$. \cite{II}
The induced interactions from multi-components $j\ge 3$ 
enhance superfluidity in such gases as well as in symmetric nuclear
matter ($\nu=4$) and color superconductivity in quark matter (8
color, 2 spin and 2-3 flavor states) \cite{long}. 

The above results apply to weakly attractive systems.
For strong attractions near the unitary limit (BCS-BEC crossover)
the induced interaction may differ. In a recent calculation\cite{Yu} 
based on the Nozieres \& Schmitt-Rink approach the induced interactions
reduce pairing by only twenty percent whereas experiments \cite{Thomas} indicate
a reduction by a factor $\sim 2-3$ compatible with an extrapolation
from the dilute limit including induced interactions \cite{long}. 

{\bf Repulsive on-site interactions} $U_{12}>0$ inhibit s-wave pairing
unless a longer range attraction is added such as a nearest
neighbor interaction $V\sum_{\langle ij\rangle}n_in_j$.  Super-exchange in the
Hubbard model generates a similar nearest neighbor (spin-spin)
interaction with coupling $J=4t^2/U_{12}$, which is believed to be responsible 
for high temperature superconductivity \cite{Anderson}.  
In the limit $U_{12}\to \infty$
extended s-wave pairing can occur but requires a strong nearest neighbor attraction
$V<-\pi^2t/2$ \cite{Micnas,Nozieres}, whereas d-wave pairing occurs naturally 
in 2D for
even weak nearest neighbor attraction. The d-wave mean field gap
equation is (see, e.g., \cite{Micnas,Nozieres})
\bea
  1 = -\frac{V}{4} \sum_{\bf q} \frac{\eta^2_{\bf q}}{2E_{\bf q}} \,,
\eea
where now $E_{\bf q}=\sqrt{\epsilon^2_{\bf q}+\Delta_d^2\eta^2_{\bf q}}$
and $\eta_{\bf q}=2[\cos q_x-\cos q_y]$.
At low filling the d-wave gap can be calculated within mean field
to leading orders in the density \cite{gos}
\bea
   \Delta_d = \frac{t}{\sqrt{n}}
\exp\left[\frac{4}{\pi n^2}\left(\frac{t}{V}+c_0+c_1n+c_2n^2\right)\right] \,.
\eea
The higher order corrections in density are:
$c_0=4/\pi-1\simeq0.27$, $c_1=\pi/2-1\simeq0.57$ and $c_2\simeq0.09$. 

At half filling we can calculate the d-wave gap within mean field as above
if correlations can be ignored, which requires that the on-site interaction is small.
To leading logarithmic orders the d-wave pairing gap is
\bea \label{TvH}
   \Delta_d = \frac{8}{e^2} t\exp\left[-\pi\sqrt{t/|V|}\right] \,,\quad n=1\,.
\eea
It is a coincidence that the prefactor $8/e^2$ is the same as in
Eq. (\ref{G3D}). Correlations are, however,
expected to suppress the d-wave mean field gap of Eq. (\ref{TvH})
near half filling when the on-site repulsion is strong.

The influence of induced interactions on anisotropic pairing
may generally be calculated from the angular
dependence of the screening function $\bar{\chi}_l\equiv\langle
\chi({\bf p}/2k_F)\cos(l\theta)\rangle$ for 2D multipoles
$l=0,1,2,..$, where $\theta$ is the angle between ${\bf p}_1$ and
${\bf p}_2$ (see \cite{Rav} for the 3D case).  For a two-component 3D
system at low density we find for the p-wave
$\chi^{3D}_1(\mu_1,\mu_1)=\log(e/4)^{1/5}\simeq-0.08$, implying that
the p-wave induced interactions are attractive and enhance p-wave
pairing by a factor $(4/e)^{1/5}\simeq1.08$. Although small, it may be
important in low density neutron matter \cite{Rav}. In 2D at low density,
however, the screening function was constant for two components as well
as multi-components whenever $\mu_j>\mu_1$ and since the Fermi surface
is circular symmetric all higher multipoles than the s-wave
vanish. Consequently, d-wave pairing is unaffected by induced
interactions in this limit and therefore the attractive super-exchange
interactions in the spin-exchange channel are not suppressed by
induced interactions.  However, adding multi-components or approaching
half filling the screening
function is no longer symmetric and one can enhance or suppress d-wave
superfluidity depending on the sign of the interactions.  Near half
filling the induced interactions are generally large as for s-wave and
therefore 2D d-wave superfluidity range from strongly suppressed
for a two-component system to possibly strongly enhanced in a
multi-component system.

{\bf The critical temperature} $T_c$ can in 3D be calculated from the
gap equation (\ref{sgap}) whereas in 2D a
Berezinskii-Kosterlitz-Thouless transition occurs \cite{Scalettar} at
a lower temperature $T_{BKT}$. At low density the transition
temperature is in 3D generally proportional to the gap with the same
prefactor $T_c/\Delta=e^E/\pi\simeq0.567..$ ($E=0.577..$ is Euler's
constant) for both s- and d-wave superfluidity with and without
induced interactions \cite{PS}. In 2D a mean field critical
temperature $T_{MF}>T_{BKT}$ \cite{Scalettar}, can be calculated from
the gap equation as above and we find that the same ratio applies even
near half filling where logarithmic singularities appear. This implies
that induced interactions change the mean field critical temperatures
by the same factor as the gap of Eq. (\ref{D}).

{\bf In summary}, ultracold Fermi atoms in traps and optical lattices
provide important systems where we can study and test superfluidity
and many-body effects in detail for multi-component systems.  2D and
3D optical lattices have different pairing properties as calculated
above for gaps, critical temperatures and induced interactions. The
latter can have strong effects on s- and d-wave pairing especially
near half filling, which affects the phases and shell structures near
the Mott insulator plateaus in traps and confined optical lattices.

\vspace{-.4cm}


\end{document}